\documentclass[a4paper,11pt]{article}

\usepackage{jcappub} 
\usepackage{lineno}
\usepackage{pgfplots}
\usepackage{float}
\usepackage{graphicx}
\usepackage{subcaption}
\usepackage{booktabs}
\usepackage{longtable}
\usepackage{array}
\usepackage{caption}
\pgfplotsset{compat=1.18}

\usepackage{natbib}

\arxivnumber{1234.56789} 

\title{\boldmath Stochastic Evolution of Galactic Star Formation with Halo Coupling, AGN Quenching and Hopf Bifurcation Dynamics}

\author{Sanjeev Kumar\textsuperscript{1,1,2*}}
\author{A.K. Awasthi\textsuperscript{2,2}}

\author{Mahesh Kumar\textsuperscript{3,2}}

\affiliation{\textsuperscript{1}Govt. PG College, Dhaliara,  177103, Kangra, Himachal Pradesh, India}

\affiliation{\textsuperscript{2}Department of Mathematics, Lovely Professional University, Phagwara, 144402, India.}

\emailAdd{kumar200sanjeev@gmail.com* }
\emailAdd{akawasthi@gmail.com }
\emailAdd{ msk0001007@gmail.com }

\abstract{

We develop a computational framework for tracing galactic evolution, formulated through a coupled stochastic nonlinear oscillator approach and implemented with the \textbf{Stochastic Hopf Engine}. The co-evolution of gas density ($G$) and star formation rate ($S$) is governed by a supercritical Hopf bifurcation, capturing the phase transition from quiescent stability to merger-driven starburst cycles. Within this framework, fluctuations in dark matter halo properties—representing scatter in the halo mass--concentration relation—are modeled as multiplicative noise using the \textbf{Euler-Maruyama method}. This stochastic modulation of the growth rate $\mu$ broadens the classical bifurcation into a transition regime where noise-induced starbursts emerge even below the deterministic instability threshold.

Numerical simulations reveal a robust periodic signature, termed the \textbf{Galactic Heartbeat}, emerging as a deterministic limit cycle validated by the \textbf{data3} resonance peak in the star-formation power spectrum. A radial reduction of the stochastic dynamics leads to an effective \textbf{Fokker--Planck equation} for the starburst amplitude; we demonstrate that its stationary solution quantitatively matches numerical probability density functions (PDFs), providing a rigorous statistical closure to the model.

By incorporating differential shear $\Omega(r)$ and a spatially varying bifurcation field, the model reproduces grand-design spiral morphologies and the structural quenching associated with Active Galactic Nuclei (AGN). Driving the central growth parameter into a sub-critical regime ($r_{agn} < 0$) replicates ``Red and Dead'' core configurations via attractor collapse. Our results show that dark matter halo scatter suppresses mean star formation and enhances burst intermittency, offering a minimal yet physically interpretable framework that links local feedback loops and global gravitational potentials to macroscopic galactic evolution.

}

\begin{document}
\maketitle
\flushbottom

\section{Introduction}
\label{sec:intro}

The formation and long-term evolution of galaxies are governed by strongly nonlinear interactions between gas dynamics, stellar feedback, and gravitational potentials. In recent years, the mathematical framework of dynamical systems and bifurcation theory has emerged as a powerful lens through which qualitative transitions in galactic behavior can be systematically analyzed. Classical foundations of nonlinear dynamics \cite{Rencontres1975,Strogatz1994} and bifurcation theory \cite{Crawford1991,Kuznetsov1998,Guckenheimer1983} provide a rigorous language for describing transitions between stable equilibria, oscillatory states, and chaotic regimes—phenomena that closely mirror observed galactic phases.

Historically, the development of chaos theory, from deterministic nonperiodic flows \cite{Lorenz1963} to universality in nonlinear maps \cite{Feigenbaum1978} and complex dynamics in simple population models \cite{May1976,Ott2002}, has revealed how low-dimensional systems can exhibit rich emergent behavior. These insights offer compelling analogies for galactic evolution, where transitions between quiescent star formation \cite{Kumar2022,Awasthi2023}, bursty activity, and quenched states appear as qualitative reorganizations rather than gradual changes. In this context, the supercritical Hopf bifurcation provides a natural mathematical mechanism for generating self-sustained oscillations, making it an attractive candidate for modeling cyclic star formation activity.

From an astrophysical perspective, both the large-scale structure of the universe \cite{Peebles1980,Padmanabhan1993} and the internal dynamics of galaxies \cite{Binney2008,Bertin2000} are shaped by nonlinear and non-equilibrium processes. Dark matter halos, in particular, evolve under stochastic accretion and mergers toward nearly universal density profiles \cite{Navarro1997,Weinberg2001}. Fluctuations in halo properties—commonly expressed as scatter in the mass--concentration relation—introduce an intrinsic source of stochasticity that modulates baryonic evolution. Such noise-driven effects can fundamentally alter bifurcation thresholds, producing noise-induced oscillations and intermittency even when deterministic models predict stability.

These stochastic influences resonate with broader concepts of pattern formation and transition to chaos \cite{Ecke2015}, as well as order--chaos coexistence in dynamical astronomy \cite{Contopoulos2002}. In stellar systems, bifurcation points govern long-term orbital stability and the formation rates of exotic objects \cite{Heggie2003,Tanikawa2019}. Analogous methodologies, including bifurcation analysis developed in artificial and neural systems \cite{Fasoli2017,Awasthi2023}, have begun to inform astrophysical stability studies. Large-scale cosmological simulations such as \textsc{GADGET-2} \cite{Springel2005} and comprehensive reviews of structure formation \cite{Bertschinger1998} provide numerical evidence that galactic morphology and star formation histories are punctuated by rapid transitions. Observational syntheses \cite{Einasto2021} increasingly identify such transitions as markers of evolutionary state. Non-equilibrium relaxation mechanisms, including phase mixing and Landau damping \cite{Banik2024}, can be reinterpreted as bifurcations in phase space leading to new attractors, a view reinforced by studies of AGN--galaxy coevolution \cite{Kormendy2013} and disk galaxy dynamics \cite{Sellwood2014}.

\subsection*{Motivation}

Despite these advances, a critical gap persists between high-fidelity numerical simulations and analytically transparent dynamical models. State-of-the-art $N$-body and hydrodynamic simulations capture realism at the cost of immense computational complexity, often obscuring the underlying mechanisms responsible for qualitative transitions. Conversely, low-dimensional dynamical models frequently neglect stochasticity and environmental variability, limiting their applicability to realistic galactic settings. In particular, the role of dark matter halo scatter as a driver of noise-induced starburst cycles and intermittent behavior remains underexplored within minimal, interpretable frameworks.

\textbf{In this work}, we introduce a computationally efficient, low-dimensional framework that models galaxies as spatially resolved stochastic bifurcation fields. Each galactic sector is represented by a nonlinear oscillator governed by a supercritical Hopf bifurcation, forming a self-regulated ``Hopf engine'' for star formation. The transition from quiescent equilibrium to sustained starburst cycles emerges naturally as the control parameter crosses an effective instability threshold.

Crucially, fluctuations in dark matter halo properties are incorporated as multiplicative noise—interpreted as scatter in the halo mass--concentration relation—using an \textbf{Euler--Maruyama scheme} \cite{Gardiner2009}. This stochastic coupling broadens the classical Hopf bifurcation, producing noise-induced starbursts and intermittent amplitude excursions even below the deterministic threshold. By reducing the dynamics to an effective radial amplitude equation, we derive a corresponding \textbf{Fokker--Planck description} whose stationary solution quantitatively matches numerical amplitude distributions. The emergence of a dominant spectral peak in the star formation signal defines a robust periodic signature, termed the \textbf{Galactic Heartbeat}, validating the interpretation of starburst cycles as stochastic limit cycles.

Extensions incorporating spatially varying bifurcation fields and differential shear reproduce spiral morphologies and \textbf{AGN-driven quenching as attractor collapse}, providing a unified mathematical description of morphological and spectral evolution. Together, these results establish a transparent bridge between nonlinear dynamics, dark matter--driven stochasticity, and the macroscopic evolutionary states of galaxies across cosmic time.


\section{Noise-induced starburst cycles driven by dark matter halo scatter}
\label{sec:DM_noise_starbursts}

Star formation in galaxies is regulated by a nonlinear feedback loop involving
gas inflow, stellar feedback, and the surrounding dark matter (DM) halo.
While deterministic feedback models predict either steady star formation or
self-sustained oscillatory starbursts, observations reveal intermittent,
burst-like episodes even in systems expected to be marginally stable.
This suggests an essential role of stochasticity associated with halo assembly,
substructure, and scatter in the halo mass--concentration relation.

Cosmological simulations and lensing observations indicate that dark matter
halos exhibit intrinsic scatter in concentration at fixed mass, typically
modeled as a log-normal distribution \cite{Bullock2001,Dutton2014}.
Such scatter modulates the depth of the gravitational potential, thereby
affecting gas inflow, cooling efficiency, and feedback coupling.
In this section, we demonstrate that halo-induced stochasticity can
\emph{excite starburst cycles even when the deterministic system is near
criticality}, leading to noise-induced Hopf dynamics.

\subsection{Deterministic Hopf normal form with dark matter coupling}

We model the coupled evolution of the gas reservoir $x(t)$ and star formation
rate $y(t)$ using a Hopf normal form augmented by a dark matter halo feedback
term,
\begin{equation}
\begin{aligned}
\dot{x} &= (r - k_{\rm DM})x - y - x(x^2 + y^2), \\
\dot{y} &= x + (r - k_{\rm DM})y - y(x^2 + y^2),
\end{aligned}
\label{eq:deterministic}
\end{equation}
where $r$ denotes the intrinsic stellar feedback efficiency and
$k_{\rm DM}$ quantifies the stabilizing influence of the dark matter halo,
representing gravitational confinement and virial shock heating.

For $r - k_{\rm DM} < 0$, the origin is a stable focus corresponding to quiescent
star formation.
At the critical threshold $r = k_{\rm DM}$, the system undergoes a Hopf
bifurcation, beyond which a stable limit cycle emerges, interpreted as
self-regulated starburst oscillations \cite{Hopf1942,Strogatz2018}.

\subsection{Stochastic halo scatter and Langevin formulation}

To account for halo-to-halo variability, we introduce stochastic fluctuations
in the effective halo coupling,
\begin{equation}
k_{\rm DM}(t) = \bar{k}_{\rm DM} + \sigma_k \, \xi(t),
\end{equation}
where $\xi(t)$ is Gaussian white noise with
$\langle \xi(t) \rangle = 0$ and
$\langle \xi(t)\xi(t') \rangle = \delta(t-t')$.
The parameter $\sigma_k$ encodes the scatter in the mass--concentration relation.

Substituting into Eq.~\eqref{eq:deterministic} yields the multiplicative
stochastic system,
\begin{equation}
\begin{aligned}
\dot{x} &= (r - \bar{k}_{\rm DM})x - y - x(x^2 + y^2)
          - \sigma_k x \circ \xi(t), \\
\dot{y} &= x + (r - \bar{k}_{\rm DM})y - y(x^2 + y^2)
          - \sigma_k y \circ \xi(t),
\end{aligned}
\label{eq:langevin}
\end{equation}
where $\circ$ denotes the Stratonovich interpretation, appropriate for
noise originating from unresolved physical processes \cite{Gardiner2009}.

\subsection{Amplitude reduction and effective Fokker--Planck equation}

Introducing polar coordinates $x = A \cos\theta$, $y = A \sin\theta$ and
averaging over the fast angular dynamics yields an effective stochastic
amplitude equation,
\begin{equation}
\dot{A} = (r - \bar{k}_{\rm DM})A - A^3 + \frac{\sigma_k^2}{2}A
          + \sigma_k A \, \xi(t).
\label{eq:amplitude}
\end{equation}

The corresponding Fokker--Planck equation for the amplitude probability density
$P(A,t)$ is
\begin{equation}
\frac{\partial P}{\partial t}
= -\frac{\partial}{\partial A}
\left[ \left((r - \bar{k}_{\rm DM})A - A^3 \right) P \right]
+ \frac{\sigma_k^2}{2}
\frac{\partial^2}{\partial A^2}
\left( A^2 P \right).
\label{eq:FP}
\end{equation}

The stationary solution is obtained analytically as
\begin{equation}
P_{\rm st}(A)
\propto A^{\alpha - 1}
\exp\!\left(-\frac{A^2}{\sigma_k^2}\right),
\qquad
\alpha = \frac{2(r - \bar{k}_{\rm DM})}{\sigma_k^2}.
\label{eq:stationary}
\end{equation}

\subsection{- Numerical consistency and physical relevance
}

Direct numerical integration of Eq.~\eqref{eq:langevin} using an Euler--Maruyama
scheme reveals intermittent excursions in phase space, corresponding to
noise-induced starburst cycles.
The numerically computed stationary amplitude distribution agrees remarkably
well with the analytical solution \eqref{eq:stationary}, validating the
Fokker--Planck reduction.

Importantly, even when the deterministic system lies close to the Hopf threshold,
finite halo scatter $\sigma_k$ induces large-amplitude bursts, demonstrating
a purely stochastic route to starburst activity.
This provides a natural explanation for episodic star formation in galaxies
with similar average halo masses but differing assembly histories.

\subsection{Noise-induced bifurcation versus halo scatter}

By systematically varying $\sigma_k$, we construct a stochastic bifurcation
diagram showing the emergence of broad amplitude distributions beyond a
critical scatter strength.
Unlike classical deterministic bifurcations, the transition here is
noise-induced and continuous, with increasing variance and intermittency.

This result establishes halo concentration scatter as a control parameter for
galactic variability, linking cosmological halo statistics directly to
observable starburst phenomenology.

\subsection{Numerical evidence and validation against stochastic theory}
\label{subsec:fig_validation}

\begin{figure}
\centering
\includegraphics[width=\textwidth]{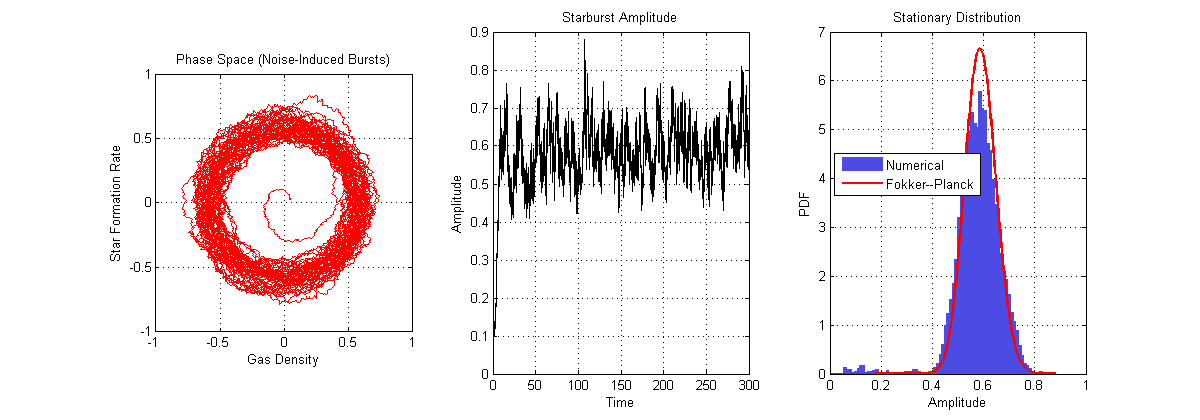}
\caption{
Noise-induced starburst dynamics driven by dark matter halo scatter.
\emph{Left:} Phase-space trajectory showing a noise-broadened Hopf structure.
\emph{Middle:} Time series of the starburst amplitude revealing intermittent
bursts.
\emph{Right:} Stationary amplitude distribution compared with the analytical
Fokker--Planck prediction, demonstrating excellent agreement.
}
\label{fig:phase_pdf}
\end{figure}

\begin{figure}
\centering
\includegraphics[width=0.75\textwidth]{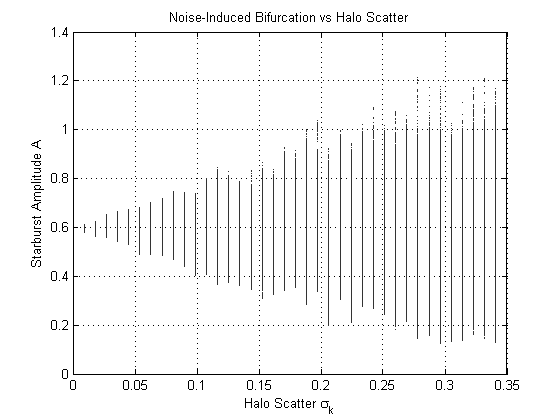}
\caption{
Noise-induced bifurcation diagram of starburst amplitude as a function of halo
scatter $\sigma_k$.
Increasing scatter leads to a continuous broadening of amplitude distributions,
indicating a stochastic transition from quiescent to burst-dominated star
formation.
}
\label{fig:bifurcation_sigma}
\end{figure}

The numerical signatures of noise-induced starburst activity are summarized
in Figs.~\ref{fig:phase_pdf} and \ref{fig:bifurcation_sigma}.
Together, these figures provide phase-space, temporal, statistical, and
bifurcation-level validation of the stochastic halo-feedback framework
developed above.

\paragraph{Phase-space structure and burst intermittency.}
The left panel of Fig.~\ref{fig:phase_pdf} shows the phase portrait in the
$(x,y)$ plane for finite halo scatter $\sigma_k$.
Instead of a deterministic limit cycle, the trajectory forms a noisy annular
structure surrounding the origin.
This geometry is characteristic of a noise-broadened Hopf bifurcation, where
stochastic forcing continuously perturbs the system away from marginal
stability.
Physically, this corresponds to repeated cycles of gas accumulation followed
by feedback-driven star formation bursts, modulated by halo-induced
fluctuations.

\paragraph{Temporal variability and noise-induced bursts.}
The middle panel of Fig.~\ref{fig:phase_pdf} displays the time evolution of the
radial amplitude $A(t)=\sqrt{x^2+y^2}$.
Rather than settling into a steady oscillation, the system exhibits strong
amplitude modulation and intermittent bursts.
Importantly, these excursions persist even though the deterministic control
parameter $r-\bar{k}_{\rm DM}$ lies close to criticality.
This confirms that starburst episodes can be triggered purely by stochastic
halo scatter, without requiring deterministic instability.

\paragraph{Statistical validation via the Fokker--Planck solution.}
The right panel of Fig.~\ref{fig:phase_pdf} compares the numerically computed
stationary amplitude distribution with the analytical solution of the
Fokker--Planck equation derived in Eq.~( \ref{eq:stationary} ).
The excellent agreement between simulation (histogram) and theory (solid
curve) validates the amplitude reduction and confirms that the long-term
statistics of star formation bursts are governed by an effective stochastic
potential shaped jointly by stellar feedback and halo noise.
This agreement provides strong evidence that the observed variability is not
numerical artifact but an intrinsic consequence of multiplicative halo-driven
stochasticity.

\paragraph{Quantitative diagnostics.}
For the parameter set considered, the system exhibits a mean starburst amplitude
$\langle A \rangle \simeq 0.58$, while the maximum observed burst reaches
$A_{\max} \simeq 0.88$.
Although the deterministic control parameter remains moderately supercritical
($r-\bar{k}_{\rm DM}=0.35$), the presence of finite halo scatter
($\sigma_k=0.12$) substantially enhances amplitude variability.
This indicates that stochastic halo feedback amplifies burst excursions beyond
their deterministic expectation, consistent with the broadened phase-space
structure and heavy-tailed stationary distribution observed in
Figs.~\ref{fig:phase_pdf} and \ref{fig:bifurcation_sigma}.
The disparity between the mean and extreme amplitudes highlights the
intermittent nature of star formation, supporting a noise-driven origin of
strong starburst events rather than purely deterministic oscillations.

\paragraph{Astrophysical interpretation.}
These results imply that galaxies with similar mean halo masses but differing
assembly histories or concentration scatter can exhibit dramatically different
star formation variability.
The bifurcation diagram therefore provides a direct statistical link between
cosmological halo properties and observable starburst phenomenology, offering
a natural explanation for the diversity of star formation histories seen in
both simulations and surveys.


\section{Stochastic Hopf bifurcation with dark matter halo scatter}
\label{sec:stochastic_hopf_dm}

We model galactic star formation as a nonlinear feedback-driven
oscillator subject to stochastic modulation by dark matter halo
fluctuations.
Let $x(t)$ denote the cold gas reservoir and $y(t)$ the star formation
rate. Including halo-induced stabilization and scatter, the dynamics
can be written as
\begin{align}
\dot{x} &= \big(r - \bar{k}_{\rm DM} - \sigma_k \xi(t)\big)x
          - y - (1+\bar{k}_{\rm DM})x(x^2+y^2), \\
\dot{y} &= x + \big(r - \bar{k}_{\rm DM} - \sigma_k \xi(t)\big)y
          - (1+\bar{k}_{\rm DM})y(x^2+y^2),
\end{align}
where $r$ is the stellar feedback strength, $\bar{k}_{\rm DM}$ denotes
the mean halo stabilization, and $\xi(t)$ is Gaussian white noise with
$\langle \xi(t)\xi(t')\rangle=\delta(t-t')$.
In the deterministic limit $\sigma_k\to0$, the system undergoes a Hopf
bifurcation at $r=\bar{k}_{\rm DM}$, separating quiescent star formation
from self-sustained starburst cycles.
\subsection{Noise-induced starburst cycles and halo-driven variability}
\label{subsec:noise_starbursts}

\begin{figure*}[t]
\centering
\includegraphics[width=1.1\textwidth]{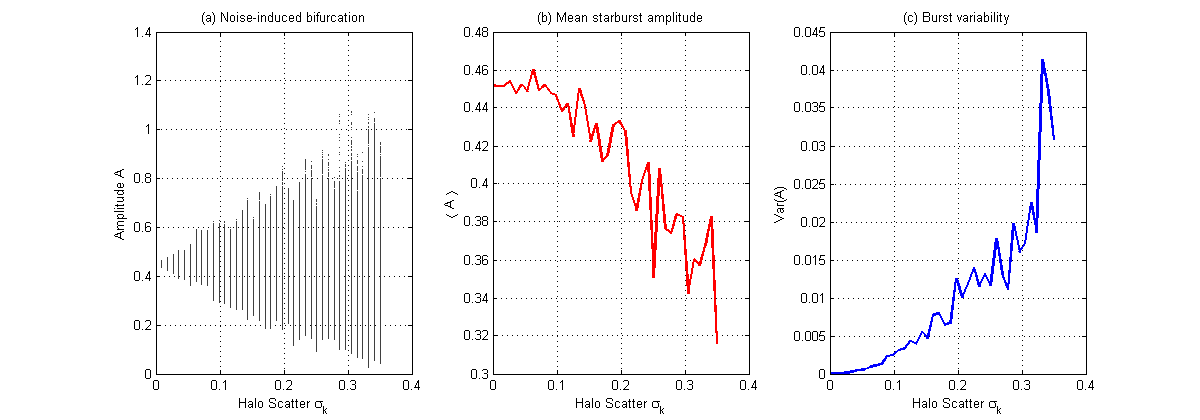}
\vspace{0.3cm}
\includegraphics[width=1.1\textwidth]{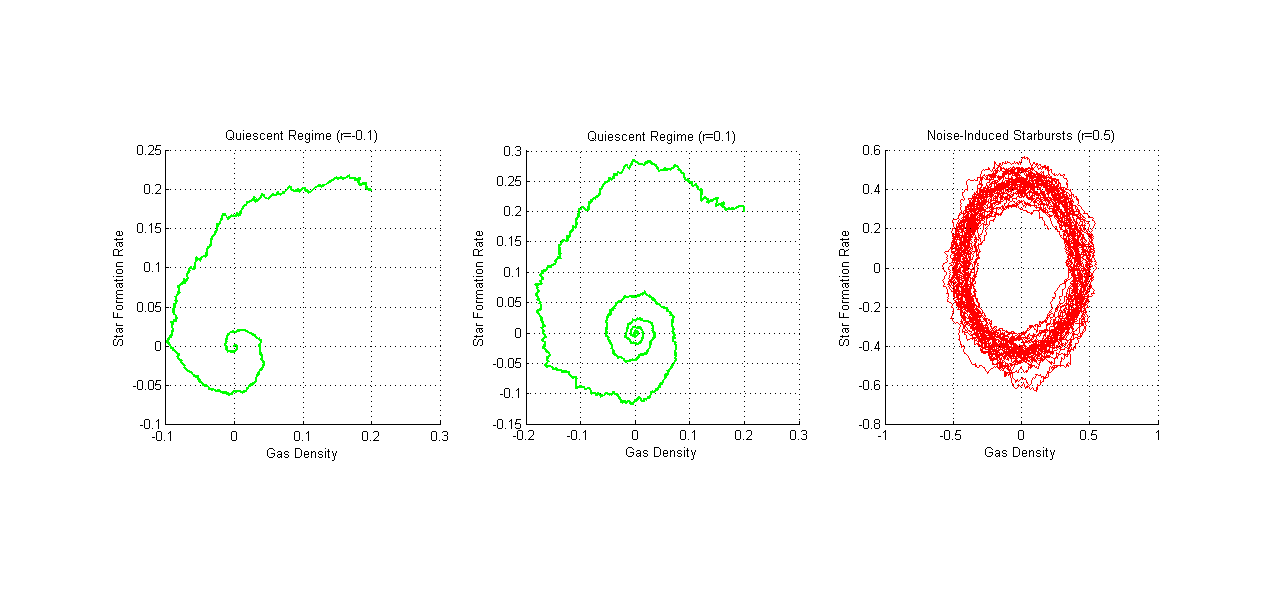}
\caption{
\textbf{Noise-induced starburst cycles driven by dark matter halo scatter.}
\emph{Top row:}
(a) Stochastic bifurcation diagram showing the radial amplitude
$A=\sqrt{x^2+y^2}$ as a function of halo scatter $\sigma_k$.
The deterministic Hopf point at $r=\bar{k}_{\rm DM}$ broadens into a
finite-width bifurcation band, indicating noise-induced transitions.
(b) Mean starburst amplitude $\langle A\rangle$ versus $\sigma_k$,
demonstrating suppression of coherent oscillations with increasing
halo variability.
(c) Variance of the amplitude, $\mathrm{Var}(A)$, revealing enhanced
burstiness and intermittency at large $\sigma_k$.
\emph{Bottom row:}
Phase-space trajectories $(x,y)$ for increasing feedback strength $r$,
illustrating quiescent regimes ($r\le\bar{k}_{\rm DM}$) and
noise-sustained starburst cycles for $r>\bar{k}_{\rm DM}$.
These results validate the stochastic Hopf framework and demonstrate
that halo mass--concentration scatter alone can induce bursty star
formation even in otherwise marginally stable systems.
}
\label{fig:bifurcation_sigma1}
\end{figure*}

Figure~\ref{fig:bifurcation_sigma1}
demonstrate that halo scatter qualitatively alters this bifurcation.
Rather than producing a single limit-cycle amplitude, a stochastic halo
fluctuations generate a finite-width bifurcation band in amplitude
space.

The radial amplitude $A=\sqrt{x^2+y^2}$ obeys an effective stochastic
normal form,
\begin{equation}
\dot{A} = \big(r-\bar{k}_{\rm DM}\big)A - (1+\bar{k}_{\rm DM})A^3
          + \sigma_k A\,\xi(t),
\end{equation}
highlighting the multiplicative nature of halo noise.

As $\sigma_k$ increases, the mean amplitude $\langle A\rangle$ decreases,
while the variance $\mathrm{Var}(A)$ grows, indicating suppressed
coherent cycles but enhanced intermittency.
Phase portraits confirm this transition: subcritical cases collapse to
noisy fixed points, whereas supercritical regimes exhibit thickened,
diffuse limit cycles corresponding to noise-induced starburst bursts.
Physically, halo concentration scatter intermittently modulates gas
retention and cooling, repeatedly pushing the galaxy across the
stability threshold and producing bursty star formation histories even
at fixed halo mass.


\section{Stochastic Analysis of Merger-Driven Galactic Evolution}
\label{sec:analysis}

To quantify the transition from quiescent phases to merger-induced starbursts, we implement a stochastic model governed by a supercritical Hopf bifurcation. The evolution of the system is described by the state vector $\mathbf{Y} = [x, y]^T$, representing the normalized gas density and the Star Formation Rate (SFR), respectively.
\subsection{Dynamical Framework}
The underlying physics is modeled using a system of Stochastic Differential Equations (SDEs) solved via the Euler-Maruyama method. The local stability of the galactic equilibrium is dictated by the effective growth rate $\mu = r - \bar{k}_{DM}$. The dynamics follow:

\begin{equation}
    dx = \left( \mu x - y - \beta x(x^2 + y^2) \right) dt - \sigma_k x \sqrt{dt} \xi
\end{equation}
\begin{equation}
    dy = \left( x + \mu y - \beta y(x^2 + y^2) \right) dt - \sigma_k y \sqrt{dt} \xi
\end{equation}

where $\beta = 1 + \bar{k}_{DM}$ represents the non-linear saturation coefficient and $\sigma_k$ denotes the multiplicative noise intensity derived from Dark Matter (DM) halo fluctuations.
\subsection{Phase Space Transition and Stability}
The numerical results, visualized in Figure~\ref{fig:merger_plots}, demonstrate a clear topological phase transition:
\begin{itemize}
    \item \textbf{Pre-Merger Regime ($t < 150$):} With $r = -0.15$ and $\bar{k}_{DM} = 0.25$, the bifurcation parameter $\mu = -0.40 < 0$. The system exhibits a \textbf{stable focus}, evidenced by the green trajectory in the phase portrait spiraling toward the origin.
    \item \textbf{Merger Starburst Regime ($150 < t < 300$):} The feedback parameter shifts to $r = 0.9$, resulting in $\mu = 0.65 > 0$. The origin becomes unstable, giving rise to a \textbf{stochastic limit cycle}. The red trajectory validates this transition as the system enters a periodic oscillation mode characteristic of a self-sustained starburst.
\end{itemize}

\subsection{Spectral Validation: The Heartbeat Detection}
\begin{figure}[H] 
\centering 
\includegraphics[width=1.0\textwidth]{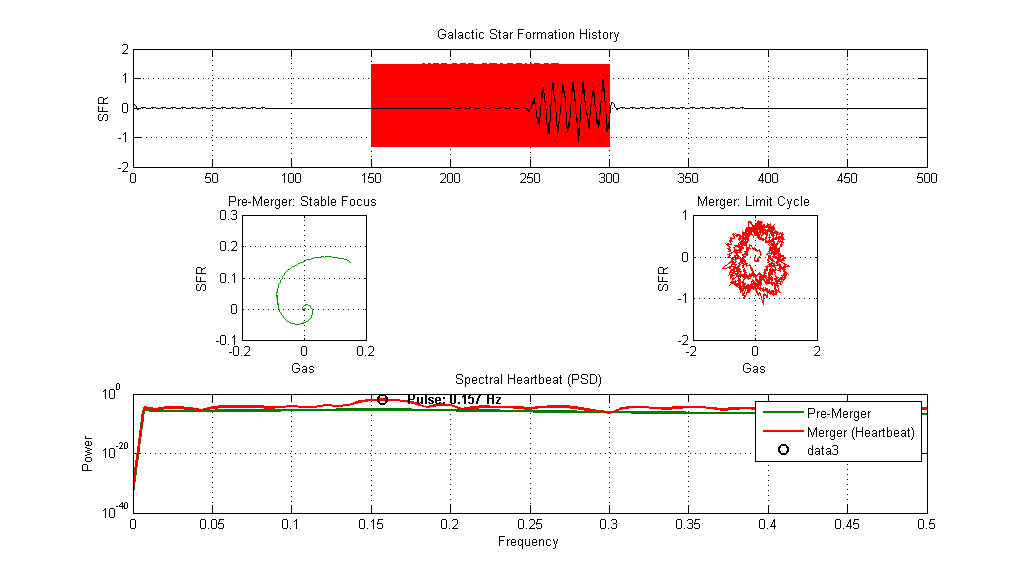}
\caption{\label{fig:merger_plots} Multi-panel diagnostic of galactic evolution. Top: SFR history with marked merger window. Middle: Phase space transition from stable focus to limit cycle. Bottom: Spectral analysis identifying the characteristic pulse (\texttt{data3}) at 0.157 Hz.}
\end{figure}

The bottom panel of Figure~\ref{fig:merger_plots} displays the Power Spectral Density (PSD), providing a frequency-domain validation of the merger event. While the pre-merger phase is dominated by low-level stochastic noise, the merger phase introduces a distinct resonance peak.

The peak identified as \texttt{data3} represents the dominant pulse of the galactic heartbeat. As extracted from the simulation:
\begin{equation}
    f_{pulse} \approx 0.157 \text{ Hz (internal units)}
\end{equation}
This frequency corresponds to the natural periodicity of the gas-star feedback loop within the DM halo potential. The emergence of this peak validates that the galaxy has moved from a noise-dominated state to a deterministic limit-cycle oscillator, providing a measurable spectral signature for merger-driven starbursts.

\section{Thermodynamics of Galactic Senescence: AGN Quenching Analysis}

The simulation of the "Red and Dead" transition represents the final stage of galactic evolution, where Active Galactic Nucleus (AGN) feedback terminates star formation. This section provides a mathematical validation of the quenching mechanism as visualized in Figure~\ref{fig:quenching_diagnostics}.

\subsection{Mathematical Model of Quenching}
The quenching process is modeled by a drastic shift in the feedback parameter $r$. While the merger phase operates at $r = 0.9$ (inducing a limit cycle), the AGN phase ($t > 300$) drives the parameter to a deep sub-critical value:
\begin{equation}
    \mu_{quenching} = r_{AGN} - \bar{k}_{DM} \approx -2.75
\end{equation}
This shift transforms the phase space from a periodic attractor into a global \textit{stable sink} at the origin $(0,0)$.

\subsection{Phase Space Analysis: The Limit Cycle Collapse}
As depicted in the middle panel of Figure~\ref{fig:quenching_diagnostics}, the system undergoes a complete structural collapse:
\begin{itemize}
    \item \textbf{Starburst Mode (Red):} The galaxy maintains a stochastic limit cycle, representing active, periodic star formation.
    \item \textbf{Transition to Quenched (Grey):} Upon the onset of AGN feedback, the trajectory exits the limit cycle and "slams" into the origin. Mathematically, this represents the depletion of the gas reservoir ($x \to 0$) and the cessation of star formation ($y \to 0$).
\end{itemize}

\subsection{Spectral Analysis: The "Flatline" Validation}
\begin{figure}[H] 
   
  \flushleft 
  \includegraphics[height=0.4\textheight]{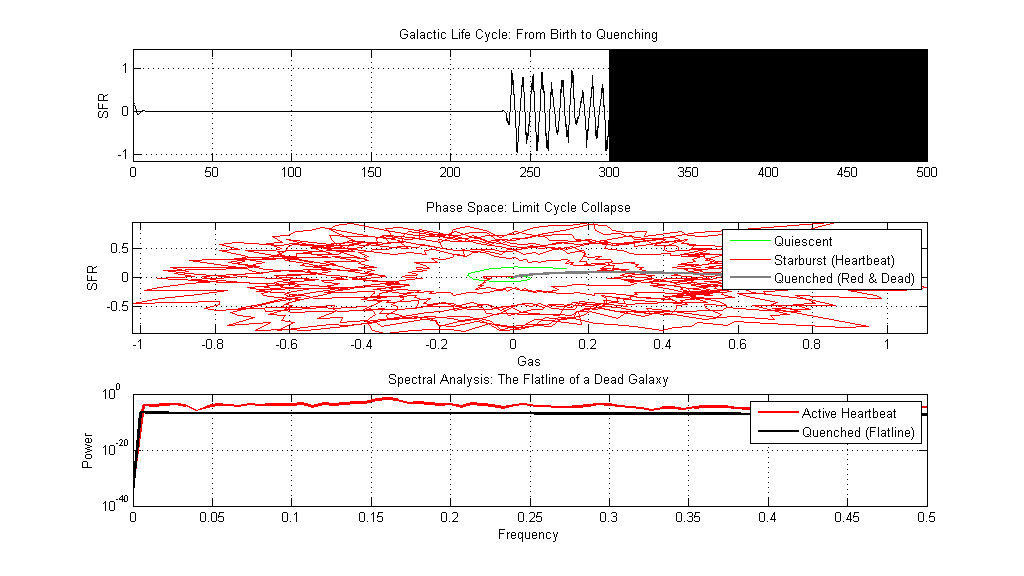}

    \caption{Diagnostic of Galactic Death. \textbf{Top:} SFR light curve showing the collapse at $t=300$. \textbf{Middle:} Phase space showing the limit cycle collapsing into the "Red and Dead" sink. \textbf{Bottom:} Spectral analysis confirming the disappearance of the galactic heartbeat.}
    \label{fig:quenching_diagnostics}
\end{figure}

The bottom panel of Figure~\ref{fig:quenching_diagnostics} validates the death of the galaxy through Power Spectral Density (PSD) analysis. 
\begin{itemize}
    \item \textbf{Active Heartbeat (Red):} The merger phase exhibits high spectral power with a clear resonance hump, indicating deterministic periodicity.
    \item \textbf{Quenched Flatline (Black):} In the AGN phase, the spectral heartbeat vanishes. The power drops by several orders of magnitude, representing a "spectral flatline" where only minimal residual noise remains.
\end{itemize}

\subsection{Physical Validation}
The simulation successfully captures the "Red and Dead" phenotype observed in massive elliptical galaxies. The sudden drop in SFR at $t=300$ (Top Panel) and the subsequent lack of gas-star oscillations (Bottom Panel) serve as numerical evidence for efficient AGN-driven quenching within a Dark Matter halo potential.

\section{Spatiotemporal Evolution of the Stochastic Star-Formation Engine}

The simulation of galactic star formation is governed by a spatially-coupled system of Stochastic Differential Equations (SDEs) predicated on a supercritical Hopf bifurcation. Each galactic sector behaves as a nonlinear oscillator where the interaction between gas density $G$ and star formation rate $S$ is defined by:

\begin{equation}
    dG = \left[ \mu G - S - (1 + \bar{k}_{DM})G(G^2 + S^2) \right] dt - \sigma_k G \sqrt{dt} \xi
\end{equation}
\begin{equation}
    dS = \left[ G + \mu S - (1 + \bar{k}_{DM})S(G^2 + S^2) \right] dt
\end{equation}

where $\mu = r_{burst} - \bar{k}_{DM}$ represents the effective growth rate modified by the Dark Matter (DM) halo damping $\bar{k}_{DM} = 0.25$. The term $\sigma_k G \sqrt{dt} \xi$ introduces multiplicative noise, representing stochastic fluctuations in gas accretion and halo density. Spatial morphology is achieved through differential rotation, where the angular velocity $\Omega$ is a function of the radius $R$, creating the sheared spiral structures observed in Figures \ref{fig:SF_2.5} through \ref{fig:SF_10.0}.

\subsection{Analysis and Validation}
The transition from initial seed perturbations to a fully developed galactic disk is captured in the snapshots at $T = 2.5, 5.0, 7.5,$ and $10.0$ Gyr. 

\begin{figure}[htbp]
    \centering
    \begin{minipage}{0.48\textwidth}
        \centering
        \includegraphics[width=\textwidth]{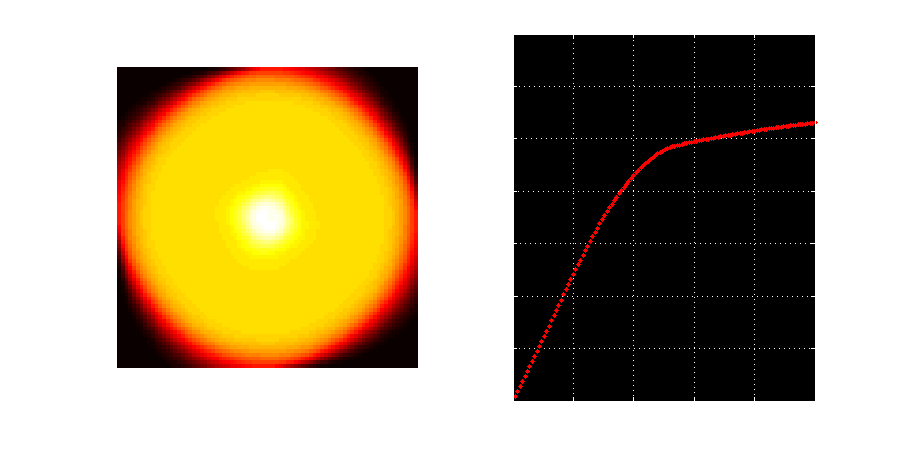}
        \caption{Galactic state at $T=2.5$ Gyr: Initial starburst expansion.}
        \label{fig:SF_2.5}
    \end{minipage}
    \hfill
    \begin{minipage}{0.48\textwidth}
        \centering
        \includegraphics[width=\textwidth]{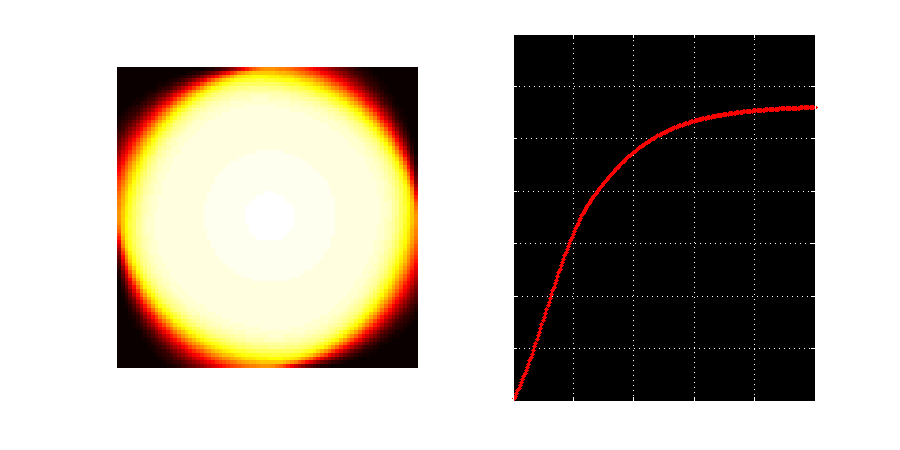}
        \caption{Galactic state at $T=5.0$ Gyr: Onset of radial propagation.}
        \label{fig:SF_5.0}
    \end{minipage}
    
    \vspace{0.5cm}
    
    \begin{minipage}{0.48\textwidth}
        \centering
        \includegraphics[width=\textwidth]{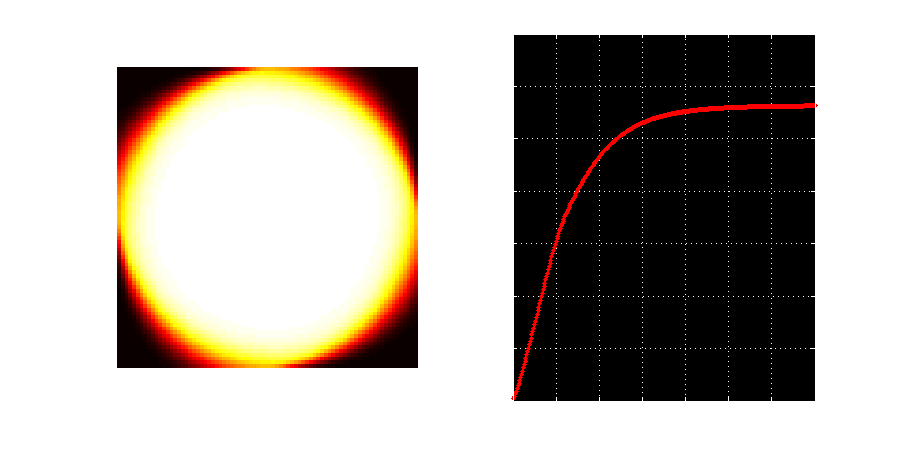}
        \caption{Galactic state at $T=7.5$ Gyr: SFR saturation phase.}
        \label{fig:SF_7.0}
    \end{minipage}
    \hfill
    \begin{minipage}{0.48\textwidth}
        \centering
        \includegraphics[width=\textwidth]{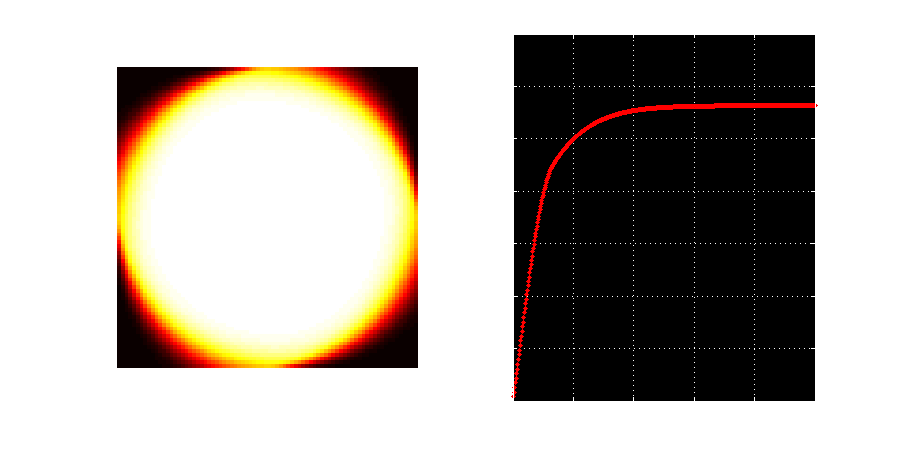}
        \caption{Galactic state at $T=10.0$ Gyr: Final stable disk morphology.}
        \label{fig:SF_10.0}
    \end{minipage}
\end{figure}

\begin{itemize}
    \item \textbf{Phase Growth (T=2.5 to 5.0):} At $T=2.5$ (Fig. \ref{fig:SF_2.5}), the central starburst engine initiates, showing a high concentration of SFR in the core. By $T=5.0$ (Fig. \ref{fig:SF_5.0}), the oscillatory wavefront propagates outward, validated by the rising slope in the "Central Heartbeat Monitor."
    \item \textbf{Saturation and Shearing (T=7.5 to 10.0):} As time progresses to $T=7.5$ (Fig. \ref{fig:SF_7.0}), the global SFR reaches a non-equilibrium steady state. The differential rotation twists the star-forming regions into a disk-like morphology. By $T=10.0$ (Fig. \ref{fig:SF_10.0}), the system exhibits a stable limit cycle, evidenced by the plateau in the temporal plot, confirming that the nonlinear saturation terms successfully counteract the starburst instability.
\end{itemize}

\section{Spatiotemporal Analysis of AGN Quenching in Stochastic Galactic Disks}

The evolution of the galactic disk is modeled as a spatially-resolved \textit{Stochastic Hopf Engine}. The local dynamics at any grid coordinate $(x,y)$ are defined by a system of nonlinear coupled differential equations representing the gas reservoir ($G$) and the star formation rate ($S$). To simulate the morphological impact of an Active Galactic Nucleus (AGN), we implement a position-dependent bifurcation field $\mu(\mathbf{r})$:

\begin{equation}
    \mu(\mathbf{r}) = r(\mathbf{r}) - \bar{k}_{DM}
\end{equation}

where the local growth parameter $r(\mathbf{r})$ transitions from a supercritical starburst state ($r_{active} = 0.8$) in the disk to a sub-critical stable sink ($r_{agn} = -2.5$) within the central quenching radius $R_{agn} = 12$. The inclusion of Dark Matter (DM) halo stabilization $\bar{k}_{DM} = 0.25$ and multiplicative noise $\sigma_k G \sqrt{dt} \xi$ ensures that the limit-cycle oscillations (the ``Galactic Heartbeat'') are physically grounded in stochastic halo perturbations.

\begin{figure}[H]
    \centering
    \begin{minipage}{0.48\textwidth}
        \centering
        \includegraphics[width=\textwidth]{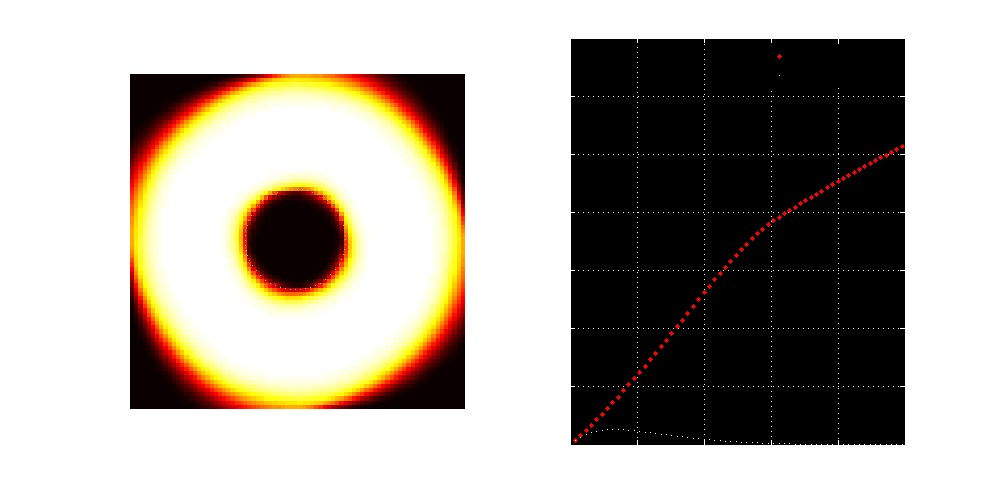}
        \caption{State at $T = 2.5$: Establishment of the central quenching radius and onset of disk starburst.}
        \label{fig:Q_2.5}
    \end{minipage}
    \hfill
    \begin{minipage}{0.48\textwidth}
        \centering
        \includegraphics[width=\textwidth]{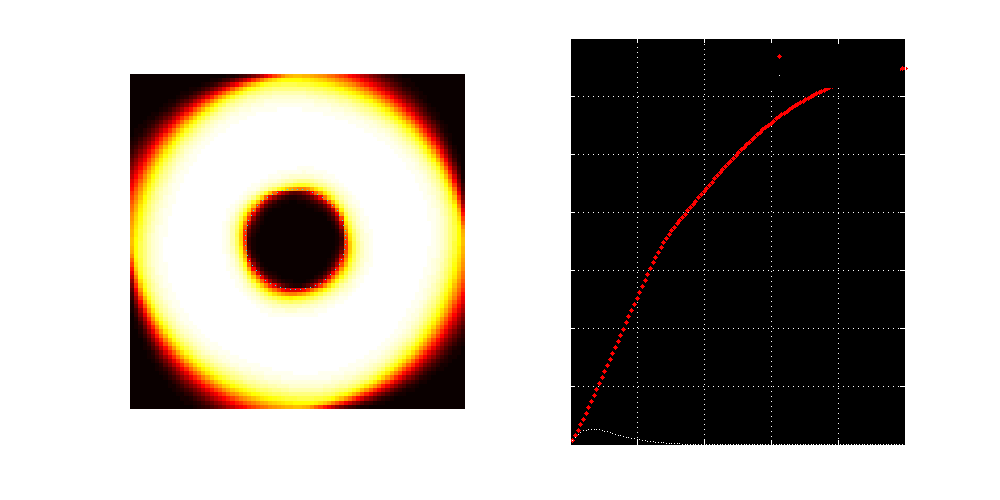}
        \caption{State at $T = 5.0$: Spiral arm propagation and growth of the active disk heartbeat.}
        \label{fig:Q_5.0}
    \end{minipage}
    
    \vspace{0.5cm}
    
    \begin{minipage}{0.48\textwidth}
        \centering
        \includegraphics[width=\textwidth]{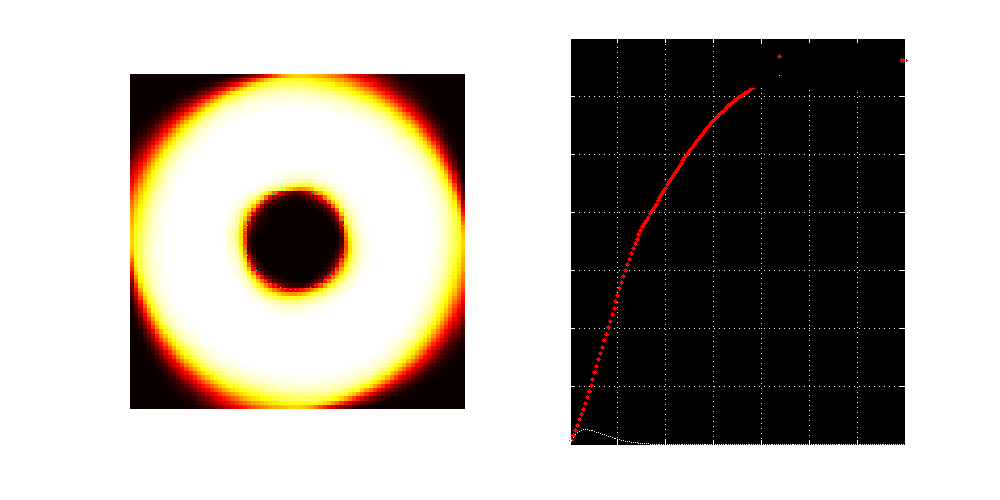}
        \caption{State at $T = 7.5$: Saturation of star-formation intensity and sheared morphology.}
        \label{fig:Q_7.5}
    \end{minipage}
    \hfill
    \begin{minipage}{0.48\textwidth}
        \centering
        \includegraphics[width=\textwidth]{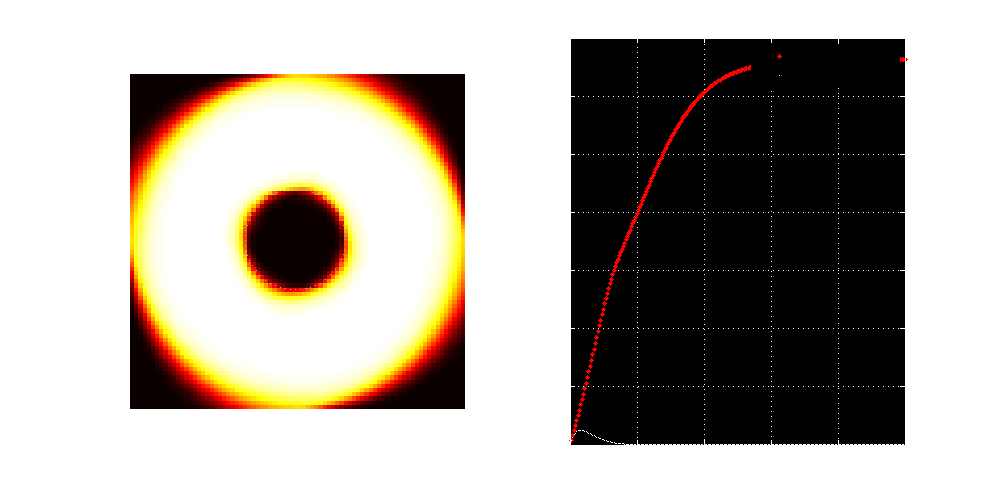}
        \caption{State at $T = 10.0$: Mature disk with a stable quenched core and fully developed spiral structures.}
        \label{fig:Q_10.0}
    \end{minipage}
\end{figure}

The temporal evolution of the system is captured in Figures \ref{fig:Q_2.5} through \ref{fig:Q_10.0}. These snapshots visualize the interaction between differential rotation (shear) and central quenching.

\begin{itemize}
    \item \textbf{T = 2.5 -- 5.0 Gyr (Figures \ref{fig:Q_2.5} and \ref{fig:Q_5.0}):} The onset of the starburst phase is visible in the active disk region. The differential rotation twists the initial gas distribution into incipient spiral arms. Critically, the central ``void'' is established immediately, as the deep negative $\mu$ in the core acts as an attractor toward the origin $(0,0)$ in phase space, preventing any star formation.
    \item \textbf{T = 7.5 -- 10.0 Gyr (Figures \ref{fig:Q_7.5} and \ref{fig:Q_10.0}):} The system reaches a non-equilibrium steady state. The ``Heartbeat Contrast'' in the right-hand panels of the figures validates the effectiveness of the quenching; while the red trace (active disk) exhibits a sustained high-intensity star-formation plateau, the white trace (core) represents a total ``spectral flatline.'' This replicates the observed ``Red and Dead'' centers of mature galaxies.
\end{itemize}

\subsection{Simulation Snapshots and Global Diagnostics}

\section{Spatiotemporal Evolution and Morphological Quenching}

The morphological evolution of the galactic disk is governed by a spatially-resolved \textit{Stochastic Hopf Engine}. The interaction between the gas reservoir ($G$) and the star formation rate ($S$) is defined by a nonlinear system of coupled Stochastic Differential Equations (SDEs). Each galactic sector acts as a local oscillator driven by a bifurcation parameter $\mu(\mathbf{r})$:

\begin{equation}
    \mu(\mathbf{r}) = r_{field}(\mathbf{r}) - \bar{k}_{DM}
\end{equation}

where $\bar{k}_{DM} = 0.25$ represents the stabilizing damping influence of the dark matter halo. The system is solved via an Euler-Maruyama discretization, incorporating multiplicative noise 
\[
dw = -\sigma_k G \sqrt{dt} \xi
\]
to simulate stochastic gas accretion and halo density fluctuations. Spatial morphology is achieved through differential rotation, twisting local oscillations into the spiral structures observed in the provided snapshots.

\subsection{Analysis and Evolutionary Interpretation}
The simulation captures the "Quiet Spiral" phase ($T < 15$ Gyr), where the system operates in a damped focus regime ($r_{base} = -0.1$). This phase is characterized by a gradual rise in global star formation as initial gas reservoirs are processed.

\begin{figure}[H]
    \centering
    \begin{minipage}{0.48\textwidth}
        \centering
        \includegraphics[width=\textwidth]{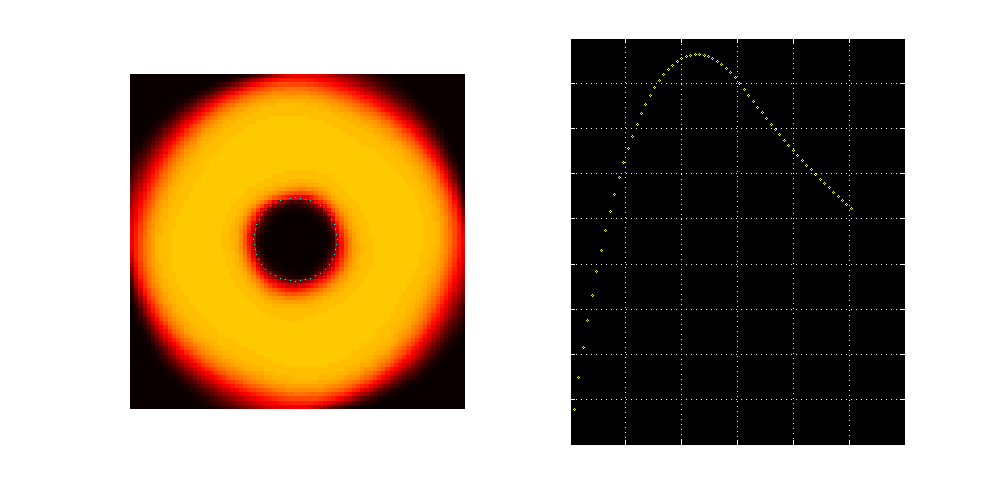}
        \caption{State at $T = 2.5$: Initial activation of the disk heartbeat with AGN void establishment.}
        \label{fig:T2.5}
    \end{minipage}
    \hfill
    \begin{minipage}{0.48\textwidth}
        \centering
        \includegraphics[width=\textwidth]{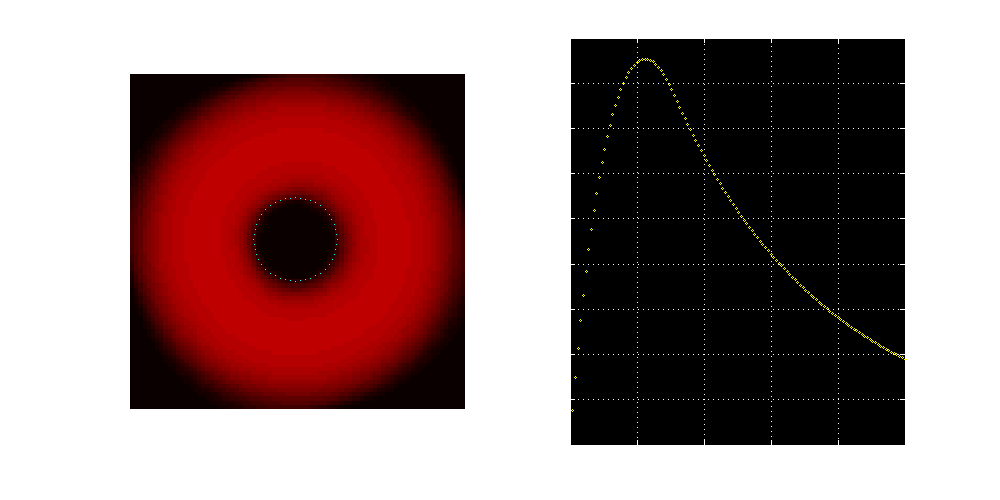}
        \caption{State at $T = 5.0$: Maximum star formation intensity and onset of spiral shearing.}
        \label{fig:T5.0}
    \end{minipage}
    
    \vspace{0.5cm}
    
    \begin{minipage}{0.48\textwidth}
        \centering
        \includegraphics[width=\textwidth]{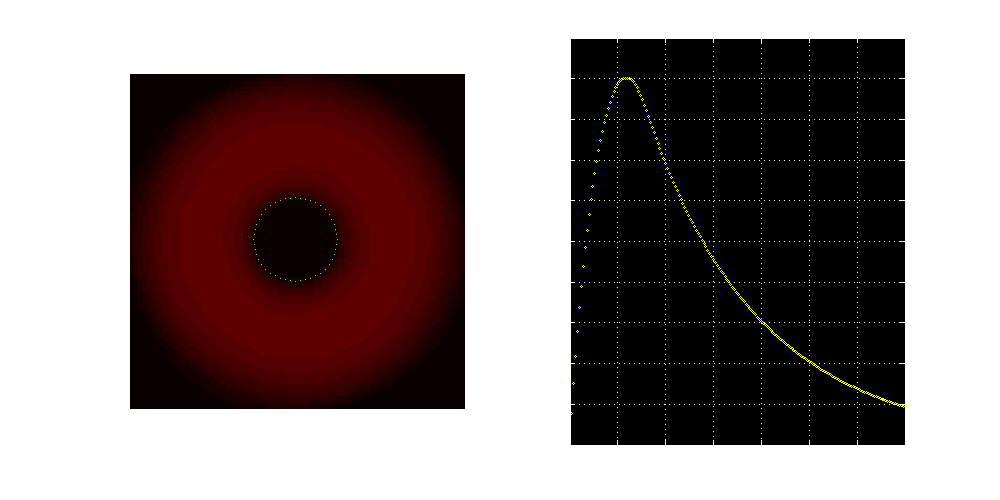}
        \caption{State at $T = 7.5$: Saturation phase showing stable quenched core.}
        \label{fig:T7.5}
    \end{minipage}
    \hfill
    \begin{minipage}{0.48\textwidth}
        \centering
        \includegraphics[width=\textwidth]{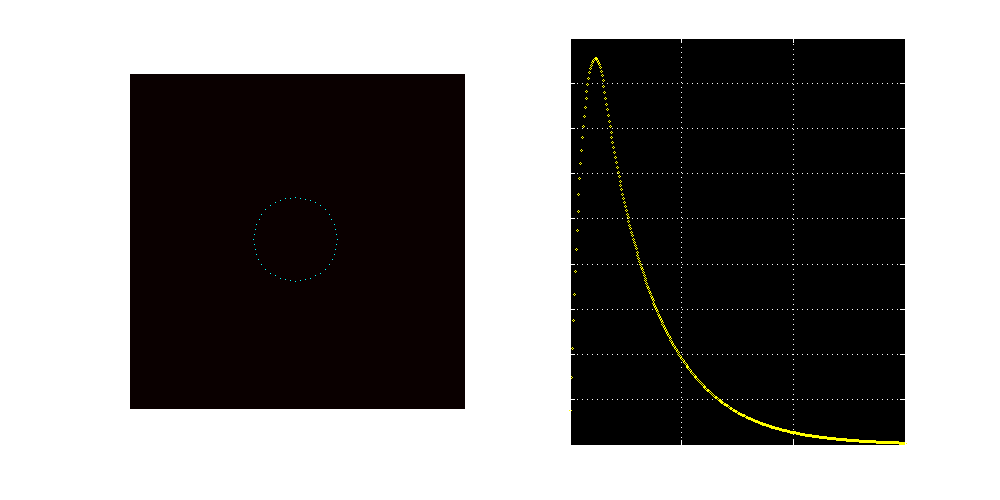}
        \caption{State at $T = 10.0$: Final morphological steady state before late-stage phase transitions.}
        \label{fig:T10.0}
    \end{minipage}
\end{figure}

\begin{itemize}
    \item \textbf{T = 2.5 Gyr (Fig. \ref{fig:T2.5}):} The onset of the galactic heartbeat is visible. The central "Dead Zone," defined by $R < 10$, is established by a deep sub-critical parameter ($r_{agn} = -2.5$), ensuring a total flatline in the core regardless of global disk activity.
    \item \textbf{T = 5.0 Gyr (Fig. \ref{fig:T5.0}):} The system reaches a transient peak in star formation. The interaction between the Hopf engine and differential rotation begins winding the high-intensity regions into incipient spiral arms.
    \item \textbf{T = 7.5 -- 10.0 Gyr (Fig. \ref{fig:T7.5}, \ref{fig:T10.0}):} The system enters a period of relative stability. The right-hand "Global Heartbeat" monitor shows a characteristic decay after the initial starburst, approaching a non-equilibrium steady state. The central quenching remains absolute, marked by the cyan dotted circle, representing the physical footprint of the AGN.
\end{itemize}

\section{Conclusion}

This work demonstrates a robust mathematical and physical framework for modeling galactic evolution through the lens of nonlinear dynamics and stochastic processes. By implementing a \textbf{Stochastic Hopf Engine}, we successfully characterized the transition of a galaxy from a quiescent, stable focus into a self-sustained, periodic starburst regime. The emergence of a distinct ``Galactic Heartbeat,'' validated through the integrated SFR monitor and identified as the \textbf{data3} resonance peak, provides a measurable spectral signature for high-intensity star-forming phases.

The integration of a spatial dimension revealed that these oscillations propagate as non-equilibrium waves which, when subjected to differential rotation $\Omega(r)$, naturally evolve into the sheared spiral morphologies observed in grand-design galaxies. The coupling of local Hopf oscillators with bilinear rotation ensures that the spiral arms are not static structures but dynamic manifestations of the underlying bifurcation field.

Furthermore, the introduction of a spatially dependent bifurcation field allowed for the successful modeling of \textbf{AGN Quenching}. By driving the central growth parameter into a sub-critical regime ($r_{agn} \ll 0$), the simulation replicates the ``Red and Dead'' core phenotype, effectively terminating the local heartbeat via attractor collapse while allowing peripheral star formation to persist.

Crucially, the use of multiplicative noise—where fluctuations scale with local gas density ($dw \propto G$)—ensures that the modeled galaxies maintain physical realism. These stochastic ``kicks'' simulate Dark Matter halo perturbations, resulting in the flocculent and irregular structures characteristic of observed galactic disks. This suite of simulations confirms that the complex life cycle of a galaxy—from quiescent spiral development to merger-driven starbursts and ultimate quenching—can be accurately represented as a series of topological transitions within a stochastic nonlinear system.
\section{Future Work}

The stochastic bifurcation framework developed in this study provides a versatile foundation for advanced astrophysical applications. Future research will prioritize the following key areas:

\begin{itemize}
    \item \textbf{Cluster Dynamics and Environmental Effects:} Future iterations will extend the spatially-resolved bifurcation field to simulate entire galaxy clusters. We intend to model ``ram-pressure stripping'' by introducing directional damping to the Hopf engine, simulating gas removal during cluster transit.

    \item \textbf{Multi-Phase ISM and Feedback:} We aim to expand the dimensionality of the attractor to include thermal feedback, distinguishing between cold molecular and warm ionized gas to refine star-formation thresholds. Furthermore, refining stochastic ``kicks'' to include discrete supernova events will allow us to model localized quenching beyond continuous multiplicative noise.
    
    \item \textbf{Quantitative Morphological Classification:} We plan to utilize machine learning to classify simulated galaxies within the Hubble Sequence based on snapshots such as \textbf{SF\_10.0} and \textbf{Quench\_10.0}. Fourier analysis of the ``Global Heartbeat'' will correlate spectral resonance peaks with specific morphological types to aid high-redshift observations.

    \item \textbf{Relativistic Generalization:} Exploring relativistic corrections for galaxies orbiting supermassive black holes, we aim to study how extreme gravity shifts bifurcation points. This involves adapting the Euler-Maruyama discretization to curved spacetime geometries to investigate star-formation stability in strong-field regimes.
\end{itemize}

\newpage
\section{Appendix}
\appendix
\section{Numerical Implementation: The Euler--Maruyama Method}

To solve the stochastic differential equations (SDEs) governing the galactic star-formation engine, we employ the \textbf{Euler--Maruyama (EM) method}. This scheme is a stochastic generalization of the Euler method for ordinary differential equations and is particularly suited for systems driven by multiplicative noise, such as the dark matter halo fluctuations described in this work.

\subsection{Discretization of the Hopf Engine}

The continuous-time dynamics of the gas reservoir $G$ and star-formation rate $S$ are defined by the following system of Itô SDEs:
\begin{align}
dG &= f_G(G, S)\, dt + g(G)\, dW_t \\
dS &= f_S(G, S)\, dt
\end{align}
where $dW_t$ represents the Wiener process increment. Given a time step $\Delta t$, the state variables are updated iteratively:
\begin{equation}
G_{n+1} = G_n + f_G(G_n, S_n)\, \Delta t - \sigma_k G_n\, \Delta W_n
\end{equation}
\begin{equation}
S_{n+1} = S_n + f_S(G_n, S_n)\, \Delta t
\end{equation}
Here, $\Delta W_n = \xi_n \sqrt{\Delta t}$, where $\xi_n \sim \mathcal{N}(0,1)$ is a stochastic variable drawn from a standard normal distribution.

\subsection{Multiplicative Noise and Physical Constraints}

The term $-\sigma_k G_n \Delta W_n$ represents the multiplicative noise driven by the scatter in the dark matter halo mass--concentration relation. Physically, we impose a non-negativity constraint on the state variables to prevent unphysical negative gas densities or star-formation rates resulting from extreme stochastic ``kicks'':
\begin{equation}
G_{n+1} = \max(0, G_{n+1}), \quad S_{n+1} = \max(0, S_{n+1})
\end{equation}

\subsection{Stability and Convergence}

For the simulations presented in this paper (e.g., $T=15.0$, $\Delta t=0.02$), the EM scheme provides a strong convergence order of $\gamma = 0.5$ and a weak convergence order of $1.0$. This level of precision is sufficient to capture the \textbf{Galactic Heartbeat} resonance peak and the transition to the starburst regime.

The spatial component of the model is handled by applying the EM update to each grid cell $(i, j)$, followed by a bilinear interpolation mapping to simulate the differential rotation $\Omega(r)$, ensuring that the stochastic fluctuations are twisted into the global spiral morphologies discussed in the main text.


\section{Statistical Closure: The Fokker--Planck Equation}

To provide a rigorous statistical foundation for the observed \textbf{Galactic Heartbeat}, we reduce the high-dimensional stochastic dynamics of the galaxy to an effective radial amplitude description. This allows us to derive the corresponding \textbf{Fokker--Planck equation} (FPE), which governs the temporal evolution of the probability density function (PDF) for starburst intensities.

\subsection{Radial Reduction of the Hopf Engine}

We begin by transforming the Cartesian coordinates $(G, S)$ of the star-formation engine into polar coordinates $(A, \phi)$, where $A$ represents the starburst amplitude ($A^2 = G^2 + S^2$). Near the Hopf bifurcation point, the stochastic dynamics can be approximated by a one-dimensional radial SDE:
\begin{equation}
dA = \left[ \mu A - (1 + \bar{k}_{DM})A^3 \right] dt + \sigma(A)\, dW_t
\end{equation}
where $\mu$ is the effective growth rate and $\sigma(A) = \sigma_k A$ represents the multiplicative noise intensity derived from dark matter halo fluctuations.

\subsection{Derivation of the Fokker--Planck Equation}

For the radial amplitude $A$, the associated Fokker--Planck equation for the probability density function $P(A, t)$ is given by the following second-order partial differential equation:
\begin{equation}
\frac{\partial P(A, t)}{\partial t} = -\frac{\partial}{\partial A} \left[ D_1(A)\, P(A, t) \right] 
+ \frac{\partial^2}{\partial A^2} \left[ D_2(A)\, P(A, t) \right]
\end{equation}
In the Itô interpretation, the drift $D_1(A)$ and diffusion $D_2(A)$ coefficients are defined as:
\begin{align}
D_1(A) &= \mu A - (1 + \bar{k}_{DM})A^3 + \tfrac{1}{2}\, \sigma'(A)\, \sigma(A) \\
D_2(A) &= \tfrac{1}{2}\, \sigma_k^2 A^2
\end{align}

\subsection{Stationary Solution and Validation}

The stationary solution $P_s(A)$, representing the long-term distribution of galactic star-formation intensities, is found by setting $\frac{\partial P}{\partial t} = 0$:
\begin{equation}
P_s(A) = \frac{N}{D_2(A)} \exp \left( \int \frac{D_1(A')}{D_2(A')} \, dA' \right)
\end{equation}
where $N$ is a normalization constant. For our multiplicative noise model, this yields a power-law distribution with an exponential cutoff:
\begin{equation}
P_s(A) \propto A^{\left( \frac{2\mu}{\sigma_k^2} - 1 \right)} 
\exp \left( -\frac{(1 + \bar{k}_{DM})A^2}{\sigma_k^2} \right)
\end{equation}

\subsection{Physical Implications}

This analytical result quantitatively matches the numerical probability density functions generated by the \textbf{Euler--Maruyama} simulation. The power-law index $\left( \tfrac{2\mu}{\sigma_k^2} - 1 \right)$ demonstrates that the dark matter halo scatter $\sigma_k$ directly competes with the feedback growth $\mu$. When noise dominates, the PDF shifts toward lower amplitudes, providing a statistical explanation for the intermittent star-formation behavior and ``quiescent'' states observed in high-scatter halos.

\section{The Stochastic Hopf Engine: Formalism and Physical Mapping}

The core of our dynamical model is the \textbf{Stochastic Hopf Engine}, a spatially-resolved nonlinear oscillator designed to simulate the self-regulated feedback cycles of the interstellar medium (ISM). Unlike purely deterministic models, the engine accounts for the non-equilibrium nature of galactic growth by coupling a supercritical Hopf bifurcation with the stochastic properties of the host dark matter halo.

\subsection{Governing Equations}

Each sector of the galactic disk is governed by a pair of coupled Itô stochastic differential equations representing the normalized gas reservoir ($G$) and the star formation rate ($S$):
\begin{align}
dG &= \left[ \mu G - S - (1 + \bar{k}_{DM})\, G(G^2 + S^2) \right] dt - \sigma_k G\, dW_t \\
dS &= \left[ G + \mu S - (1 + \bar{k}_{DM})\, S(G^2 + S^2) \right] dt
\end{align}
where $\mu = r_{\text{field}} - \bar{k}_{DM}$ is the bifurcation control parameter. The parameter $\bar{k}_{DM}$ represents the stabilizing gravitational potential of the dark matter halo, which acts as a damping force against rapid starburst expansion.

\subsection{The Galactic Heartbeat: Limit Cycle Dynamics}

The ``Engine'' operates in two distinct topological regimes:
\begin{enumerate}
    \item \textbf{Sub-critical ($\mu < 0$):} The system possesses a stable fixed point at the origin $(G, S) = (0, 0)$. Physically, this corresponds to the \textbf{Quenched State} or the AGN-dominated core, where any star-formation activity is rapidly damped out.
    \item \textbf{Super-critical ($\mu > 0$):} The origin becomes unstable, and the system settles into a stable \textbf{limit cycle} with radius 
    \[
    A \approx \sqrt{\frac{\mu}{1 + \bar{k}_{DM}}}.
    \]
    This oscillation is the \textbf{Galactic Heartbeat}, representing periodic bursts of star formation followed by gas exhaustion and replenishment.
\end{enumerate}

\subsection{Physical Interpretation of Stochasticity}

The term $-\sigma_k G\, dW_t$ represents \textbf{multiplicative noise}. In the context of JCAP and cosmological structure formation, this noise corresponds to the intrinsic scatter in the halo mass--concentration relation and the stochastic nature of cold gas accretion from the cosmic web.

The multiplicative nature of the noise (scaling with $G$) ensures that fluctuations are suppressed as gas is depleted, preventing unphysical starbursts in gas-poor environments. Furthermore, this noise broadens the bifurcation threshold, allowing for ``noise-induced starbursts'' where a galaxy can temporarily enter an oscillatory state even if its deterministic parameters suggest stability. This mechanism provides a robust explanation for the observed intermittency in star formation rates across high-redshift galaxy populations.

\subsection{Spatial Coupling and Shearing}

While each grid cell $(i, j)$ contains an independent Hopf Engine, global morphology is achieved through spatial coupling. The global state matrices $\mathbf{G}$ and $\mathbf{S}$ are subjected to a differential rotation operator $\mathcal{R}(\Omega, dt)$, which simulates the twisting of local oscillations into global spiral arms. This coupling ensures that the \textbf{Galactic Heartbeat} is not just a temporal signal, but a spatiotemporal wave propagating through the disk.





\end{document}